\documentclass[12pt]{article}

\usepackage{amsmath, amssymb, graphics, setspace}
\usepackage{amsmath}
\usepackage{amsxtra}
\usepackage{amstext}
\usepackage{amssymb}
\usepackage{latexsym}
\usepackage{dsfont} 
\usepackage{graphicx}

\newcommand{\be}{\begin{equation}}
\newcommand{\ee}{\end{equation}}
\newcommand{\bea}{\begin{eqnarray}}
\newcommand{\eea}{\end{eqnarray}}

\begin{document}

\begin{titlepage}
\begin{center}

\hfill SISSA 61/2011/EP\\

\vskip .3in \noindent


{\Large \bf{Orbifold Vortex and Super Liouville Theory}} \\

\vskip .2in

{ Jian Zhao}

\vskip .05in
{\em\small
International School of Advanced Studies (SISSA) \\via Bonomea 265, 34136 Trieste, Italy  }
\vskip .05in

\end{center}
\begin{center} {\bf ABSTRACT }
\end{center}
\begin{quotation}\noindent
We study the nonabelian vortex counting problem on $\mathbb{C}/\mathbb{Z}_p$. At first we calculate vortex partition functions on the orbifold space using localization techniques, then we find how to extract orbifold vortex partitions function from orbifold linear quiver instanton partition functions. Finally, we study the AGT like relation
between orbifold $SU(2)$ vortices and $\mathcal{N} = 1$ super Liouville theory in the mixed R/NS sector
by fixing the dictionary among parameters in the common hypergeometric functions system.

\end{quotation}
\vfill
\eject

\end{titlepage}

\section{Introduction}
New connections between two completely different theories will generate interesting discoveries on both sides. One fair example in recent years is the discovery of the duality between $\mathcal{N}=2$ quiver gauge theory and Liouville conformal field theory \cite{AGT}. In \cite{surface}, \cite{BTZ1}and \cite{BTZ2} the relations among  surface operators of $\mathcal{N}=2$ four dimensional gauge theories, degenerate fields of Liouville theory and two dimensional vortex theories are studied in detail. Recently, the AGT correspondence related to  ALE instanton counting has been studied in \cite{superLiouvilleNS}, \cite{superLiouville} and \cite{Ito}.

After \cite{nav}, non-abelian vortices became a hot area of research. The moduli space of vortices on a Riemann surface was studied in \cite{vortexmoduli} and the moduli space of vortices on $\mathbb{C}/\mathbb{Z}_p$ was studied in \cite{orbifold}. We use analogous localization techniques used in \cite{BTZ1} to calculate vortex partition functions on $\mathbb{C}/\mathbb{Z}_p$, which is a similar extension of instanton partition functions on $\mathbb{C}^2$ to $\mathbb{C}^2/\mathbb{Z}_p$ calculated in \cite{Fucito2004} and \cite{Dijkgraaf}. This similarity is expectable from the string theory point of view.

In the context of string theory, linear quiver gauge theories have a geometrical realization as the low energy effective theories of $D4$-branes intersecting with $NS5$-branes \cite{four}, and instantons can be considered as $D0$-branes inside the $D4$-branes.
By localization techniques, integrations over instanton moduli space turn into combinatoric problems associated with an arrow of two dimensional Young-tableaux and  each $D0$-brane is associated with a box in the Young-tableaux \cite{Nek}.
When mass parameters are in special values, the instanton partition functions will degenerate into simpler forms characterized by one dimensional Young-tableaux \cite{BTZ2}. We study the degeneration phenomenon of orbifold quiver instanton partition functions which not only tells us how to extract orbifold vortex partition functions from that of instantons but also gives information about surface operators of orbifold gauge theory.

After studying the relation between orbifold vortices and orbifold instantons, one is urged to study the AGT dual of orbifold vortices. It is difficult to find the conformal field theory dual of vortex partition functions directly. The trick here is that we can use four dimensional gauge theories as a bridge connecting conformal field theories and vortex theories \cite{BTZ2}. The AGT dual of correlation functions of pure NS primary fields was studied in \cite{superLiouvilleNS} and that of Whittaker vectors in the Ramond sector was studied in \cite{Ito}. However, our analysis shows that in order to find the AGT dual of orbifold vortices, it is necessary to have a complete knowledge of the AGT duality of super Liouville theory with both NS and Ramond sectors,  which worths a single paper by itself. We study the super Liouville theory dual of orbifold vortices based on known results about correlation functions of degenerate fields in Ramond sector \cite{hep-th/0202032}, \cite{Chorazkiewicz2011} and show that orbifold vortex partition functions can be identified with correlation functions of lowest degenerate states in the Ramond sector.

The organization of this paper is as follows.
In section 2 we review necessary knowledge about instanton counting on $\mathbb{C}^2/\mathbb{Z}_p$.
In section 3 we calculate vortex partition function on $\mathbb{C}/\mathbb{Z}_p$.
In section 4 we give the relation between the two classes of partition functions obtained in section 2 and section 3.
Then in section 5 we study the CFT dual of vortex partition function on $\mathbb{C}/\mathbb{Z}_p$.
Section 6 contains discussions.

\section{Instantons on $\mathbb{C}^2/\mathbb{Z}_p$}
In this section we will review how to do instanton counting for $U(N)$ linear quiver gauge theory on the orbifold space $\mathbb{C}^2/\Gamma$, where $\Gamma=\mathbb{Z}_p$. \cite{Fucito2004} is a standard reference for this topic. We use $k$ to denote the instanton number and parameters for pure instanton counting are Coulomb branch parameters \(a_{\alpha }\) where $\alpha $ runs from 1 to \(N\) and the $\Omega$-deformation
parameters, \(\epsilon _1,\epsilon _2\). Due to the orbifold  action, $a_\alpha, \epsilon_1, \epsilon_2$ have respectively discrete charges $q_\alpha, 1, -1$. Notice that discrete charges take value in \(\mathbb{Z}_p\), so two charges are the same if they are congruent modulo $p$. Since \(\mathbb{Z}_p\) commutes with the gauge groups, under this assignment of charges, the gauge groups will break in the following way:
\bea
&&U(N)\longrightarrow \prod _qU\left(n_q\right), \nonumber\\
&&n_q=\sum _{\alpha }\delta _{q,q_{\alpha }} . \nonumber
\eea
It seems that $\Gamma$ will change the fixed point structure of instanton counting drastically, but due to the fact that \(\Gamma \in U(1)^2\) of the localization torus action, fixed points are still characterized by $N$ Young tableaux of total number of boxes equals to k. Similarly the auxiliary $U(k)$ group will also break as :

\be
U(k)\longrightarrow \prod _qU\left(k_q\right) .\nonumber
\ee

As we know each box in a Young diagram represents an instanton, and the corresponding discrete charge is just $q_{\alpha }+i-j$  for a box at position $(i,j)$ of the $\alpha$-th Young diagram. So $k_q=\dim  V_q=$number of instantons with discrete charge $q$. Here $V$ and $W$ are complex linear spaces of dimension $k$ and $N$.  Then we have following linear decomposition of the Euler character of the tangent bundle of instanton moduli space  :
\bea
\chi _{\Gamma }&=&V^*\otimes V\left(T_1+T_2-1-T_1T_2\right)+W^*\otimes V+V^*\otimes W T_1T_2\nonumber\\
&=&\sum _q \left(V_q^*V_{q+1}+V_{q+1}^*V_q-V_q^*V_q T_1T_2-V_q^*V_q+W_q^*V_q+V_q^*W_q T_1T_2\right),\nonumber\\
V_q&=&\sum _{\alpha =1}^N\sum _{s\in Y_{\alpha }} T_{a_{\alpha }}T_1^{-j_s+1}T_2^{-i_s+1}\delta _{q_{\alpha }+i_s-j_s,q},\\
W_q&=&\sum _{\alpha =1}^NT_{a_{\alpha }}\delta _{q_{\alpha },q}.\nonumber
\eea

After some algebra we get
\bea
\chi _{\Gamma }^{\text{vector}}&=&-\sum _{\alpha ,\beta }^N\sum _{s\in Y_{\alpha }} \left(T_{a_{\alpha ,\beta }}T_1^{-L_{\beta }(s)}T_2^{A_{\alpha }(s)+1}+T_{a_{\beta ,\alpha }}T_1^{L_{\beta }(s)+1}T_2^{-A_{\alpha }(s)}\right)\delta _{L_{\beta }(s)+A_{\alpha }(s)+1,q_{\alpha ,\beta }}\nonumber\\
&=&-\sum _{\alpha ,\beta }^N\sum _{s\in Y_{\alpha }} T_{a_{\alpha ,\beta }}T_1^{-L_{\beta }(s)}T_2^{A_{\alpha }(s)+1}\delta _{L_{\beta }(s)+A_{\alpha }(s)+1,q_{\alpha ,\beta }}\\
& &-\sum _{\alpha ,\beta }^N\sum _{t\in Y_{\beta }} T_{a_{\alpha ,\beta }}T_1^{L_{\alpha }(s)+1}T_2^{-A_{\beta }(s)}\delta _{L_{\alpha }(s)+A_{\beta }(s)+1,q_{\beta ,\alpha }}.\nonumber
\eea

To obtain 4d instanton partition functions, we need to set  \(T_1=e^{ \epsilon _1}, T_2=e^{  \epsilon _2}, T_{a_{\alpha }}=e^{  a_{\alpha }}\) and then take the four dimensional limit.
As it is known from \cite{BTZ1} and \cite{BTZ2}, vortex partition functions lie in \(\epsilon _+=\epsilon_1+\epsilon_2=0\) limit of degenerate instanton partition functions, we will take this limit in the following:

\bea
\left(Z_{\Gamma }^{\text{vector}}\left(a,Y,{q_{\alpha}}\right)\right)^{-1}&=&
\prod _{\alpha ,\beta }^N\prod _{s\in Y_{\alpha }} \left(a_{\alpha ,\beta }+\epsilon _2\left(A_{Y_{\alpha }}(s)+1+L_{Y_{\beta }}(s)\right)\right)\delta _{A_{Y_{\alpha }}(s)+1+L_{Y_{\beta }}(s),q_{\alpha ,\beta }}
\nonumber\\
&&\prod _{t\in W_{\beta }} \left(a_{\alpha ,\beta }-\epsilon _2\left(A_{Y_{\beta }}(t)+1+L_{Y_{\alpha }}(s)\right)\right)\delta
_{A_{Y_{\beta }}(t)+1+L_{Y_{\alpha }}(t),q_{\beta ,\alpha }} .
\label{inst:vector}
\eea

Vector field contributions are in denominators of instanton partition functions, and numerators of instanton partition function will come from  hypermultiplets. For our interest lies in linear quiver gauge theories, we will only consider hypermultiplets in (anti)fundamental and bifundamental representations.  Since latter we will study $N$-node quiver gauge theory, we will take following notations:
\bea
\left\{Y_{\alpha }^{(L)}\right\}_{\alpha =1}^N &:&\text{  the Young tableaux of the $L$-th gauge factor.}\\
\nonumber\\
\left\{a_{\alpha }^{(L)}\right\}_{\alpha =1}^N &:&\text{  the Coulomb branch parameters of the L-th gauge factor.}\nonumber\\
m_i &=& \text{the i-th mass of bifundamental hypermultiplet}\\
\mu _i&=&\left\{
\begin{array}{cc}
 \text{masses of antifundamental hypermultiplets} & i\in [1,N] \nonumber\\
 \text{masses of fundamental hypermultiplets} & i\in [N+1,2N]
\end{array}
\right.\\
m_{\alpha ,\beta }^{(L)}&\text{:=}&a_{\alpha }^{(L)}-a_{\beta }^{(L+1)}-m_L\nonumber\\
\nonumber\\
\left\{q_{\alpha }^{(L)}\right\}_{\alpha =1}^N &:& \text{the discrete charges of Coulomb branch parameters of the L-th gauge factor.}\nonumber\\
q_m^{(L)} &:& \text{the discrete charge of the $L$-th bifundamental hypermultiplet. }\nonumber\\
q^{f}_{\alpha} &:& \text{the discrete charge of the $\alpha$-th fundamental hypermultiplet. }\\
q^{af}_{\alpha} &:& \text{the discrete charge of the $\alpha$-th antifundamental hypermultiplet. }\nonumber\\
Q_{\alpha ,\beta }^{(L)}&=&q_{\alpha }^{(L)}-q_{\beta }^{(L+1)}+q_m^{(L)} .\nonumber
\eea

\subsection{With bifundamental matter fields}
From the vector field contribution, we can easily obtain the contribution  from bifundamental hypermultiples:
\bea
\chi _{\Gamma }^{\text{bifund},L}=\sum _{\alpha ,\beta }^NT_{m_L}\left(\sum _{s\in Y_{\alpha }^{(L)}} T_{a_{\alpha ,\beta }^{(L,L+1)}}T_1^{-L_{Y_{\beta
}^{(L+1)}}(s)}T_2^{A_{Y_{\alpha }^{(L)}}(s)+1}\delta _{L_{Y_{\beta }^{(L+1)}}(s)+A_{Y_{\alpha }^{(L)}}(s)+1,Q_{\alpha ,\beta }^{(L,L+1)}}+\right.\nonumber\\
\left.\sum _{t\in Y_{\beta }^{(L+1)}} T_{a_{\alpha ,\beta }^{(L,L+1)}}T_1^{L_{Y_{\alpha }^{(L)}}(t)+1}T_2^{-A_{Y_{\beta }^{(L+1)}}(t)}\delta _{L_{Y_{\alpha
}^{(L)}}(t)+A_{Y_{\beta }^{(L+1)}}(t)+1,Q_{ \beta,\alpha }^{(L,L+1)}}\right).\nonumber
\eea

In \(\epsilon _+=0\) limit, the contribution to instanton partition function from the $L$-th bifundamental
hypermultiplet is:
\bea
Z_{\Gamma }^{\text{bifund},L}(a,m,Y)&=&\prod _{\alpha ,\beta }^N \prod _{s\in Y_{\alpha }^{(L)}} \left(m_{\alpha ,\beta }^{(L)}+\epsilon _2\left(L_{Y_{\beta }^{(L+1)}}(s)+A_{Y_{\alpha }^{(L)}}(s)+1\right)\right)\delta _{L_{Y_{\beta }^{(L+1)}}(s)+A_{Y_{\alpha }^{(L)}}(s)+1,Q_{\alpha ,\beta }^{(L)}}\nonumber\\
&   &\prod _{t\in Y_{\beta }^{(L+1)}} \left(m_{\alpha ,\beta }^{(L)}-\epsilon _2\left(L_{Y_{\alpha }^{(L)}}(t)+A_{Y_{\beta }^{(L+1)}}(t)+1\right)\right)\delta _{L_{Y_{\alpha }^{(L)}}(t)+A_{Y_{\beta }^{(L+1)}}(t)+1,Q_{\beta ,\alpha }^{(L)}}.\nonumber
\eea

\subsection{With fundamental matter fields}
It is easy to obtain contributions from fundamental hypermultiplets by either direct calculation or  reduction from that of bifundamental hypermultiplets. The results are:
\bea
Z_{\Gamma }^{\text{fund},q^f_{\beta}}(a,m,Y)&=&\prod _{\alpha =1}^N\prod _{\beta=1}^F\prod _{s\in Y_{\alpha }}\left(a_{\alpha }-m_{\beta}+\epsilon _1\left(i_s-1\right)+\epsilon
_2\left(j_s-1\right)+\epsilon _+\right)\delta _{j-i,q_{\alpha }-q^f_{\beta}},\nonumber\\
Z_{\Gamma }^{\text{antifund},q^{af}_{\beta}}(a,m,Y)&=&\prod _{\alpha =1}^N\prod _{f=1}^F\prod _{s\in Y_{\alpha }}\left(a_{\alpha }+m_{\beta}+\epsilon _1\left(i_s-1\right)+\epsilon
_2\left(j_s-1\right)\right)\delta _{j-i,q_{\alpha }-q^{af}_{\beta}}.\label{inst_fund}
\eea

\subsection{Different sectors}

For the $N$ node $SU(N)$ linear quiver theory on $\mathbb{C}/\mathbb{Z}_p$, we have different branches determined by discrete charges. The generic formula for a specific branch of orbifold instanton partition function is:
\bea
Z_{Quiver}\left(a,m,\left\{q_{\alpha }^{(L)}\right\};\left\{q_{\alpha }^{\text{af}}\right\};\left\{q_{\alpha }^f\right\}\right)&=&\sum _Y \prod _{\beta =1}^N z_{\beta}^{|Y^{\left(\beta\right)}|} Z_{\Gamma }^{\text{antifund},q_{\beta }^{\text{af}}}\left(a,m,Y^{(1)}\right)\\&&
Z_{\Gamma }^{\text{fund},q_{\beta }^f}\left(a,m,Y^{(N)}\right)
Z_N^{\Gamma }\left(\left\{q_{\alpha }^{(N)}\right\},Y^{(N)}\right)\nonumber\\&&
\prod _{L=1}^{N-1}Z_L^{\Gamma }\left(\left\{q_{\alpha }^{(L)}\right\},Y^{(L)}\right)Z_{L,L+1}^{\Gamma }\left(Y^L,Y^{L+1}\right),\nonumber
\label{orbifoldinst}
\eea
where
\bea
&&Z_L^{\Gamma }\left(\left\{q_{\alpha }^{(L)}\right\},Y^{(L)}\right)\text{:=}Z_{\Gamma }^{\text{vec}}\left(a^{(L)},Y^{(L)},\left\{q_{\alpha }^{(L)}\right\}\right),\nonumber\\
&&Z_{L,L+1}^{\Gamma }\left(Y^L,Y^{L+1}\right)\text{:=}Z_{\Gamma }^{\text{bifund},L}\left(a^{(L)},m_L,\left\{q_{\alpha }^{(L)}\right\},\left\{q_m^{(L)}\right\},Y^L,Y^{L+1}\right),\nonumber
\eea
and $z_{\beta}$ is the gauge coupling of the $\beta$-th gauge factor.
In general, orbifold instanton counting has two counting parameters if the first Chern class, $c_1$, of orbifold instanton moduli space is nontrivial. For simplicity we will only consider the case when $c_1=0$.

We will see later, in order to extract vortex partition functions from that of instantons, up-to the Weyl symmetry, we need  to choose the discrete charges  in the following way: \(q_{\alpha }^{(1)}-q_{\alpha }^f=\delta _{1,\alpha } \bmod p\) { }and \(q_{\alpha }^{(L)}-q_{\alpha }^{(L+1)}+q_m=\delta
_{\alpha ,L+1} \bmod p\).

\section{ Vortices on $\mathbb{C}/\mathbb{Z}_p$ }
The moduli space of orbifold vortex was studied in \cite{orbifold} using the moduli matrix method. In the following we will studying the orbifold vortex counting problem. As we know from \cite{BTZ1}, \cite{nav} the moduli space of vortices can be considered as a Lagrangian submanifold of the moduli space of instantons. Similar mechanism can be used for the orbifold case. Recall that the moduli space of vortex partition function on $\mathbb{C}$ is given by following ADHM like data:
\be
\mathcal{M}_{N,k}=\left.\left\{(B,I) \left| \left[B,B^{\dagger }\right]+I I^{\dagger }\right.=c \mathbb{I}_k\right\}\right/U(k),\nonumber
\ee

where \(B\in \text{End}(V,V), I\in  \text{Hom}(V,W)\). $V$ and $W$ are complex linear spaces of dimension $k$ and $N$ . When there is an extra \(\mathbb{Z}_p\) action,  $V$ and $W$ have further weight decomposition:
\bea
V_q&=&\sum _{\alpha =1}^N\sum _{j=1}^{k_{\alpha }}T_{a_{\alpha }}T_{\hbar }^{-i+1}\delta _{q_{\alpha }+i-1,q},\nonumber\\
W_q&=&\sum _{\alpha =1}^NT_{a_{\alpha }}\delta _{q_{\alpha },q},\nonumber\\
\chi _{\Gamma }&=&V^*\otimes V\left(T_1-1\right)+W^*\otimes V=\sum _q \left(V_q^*V_{q+1}-V_q^*V_q+W_q^*V_q\right).
\eea

A short calculation shows:
\be
\chi _{\Gamma }=\sum _{\alpha ,\beta =1}^NT_{a_{\alpha ,\beta }}\sum _{i=1}^{k_{\alpha }}T_{\hbar }^{-i+1+k_{\beta }}\delta _{-i+1+k_{\beta },q_{\alpha
,\beta }}.
\ee

So the vector field contribution is
\be
\left(Z_{\Gamma ,\text{vortex}}^{\text{vector}}(a,\hbar ;k;q)\right)^{-1}=
\prod _{\alpha ,\beta =1}^N\prod _{i=1}^{k_{\alpha }}\left(a_{\alpha ,\beta }+\hbar \left(k_{\beta
}+1-i\right)\right)\delta _{-i+1+k_{\beta },q_{\alpha ,\beta }}
.
\ee

Similarly, we get contributions from matter fields in fundamental representation:

\bea
Z_{\Gamma ,\text{vortex}}^{\text{fund},q^f_{\beta}}(a,m,\hbar ;k)&=&\prod _{\alpha =1}^N\prod _{\beta=1}^F\prod _{i=1}^{k_{\alpha }}\left(a_{\alpha }-m_\beta+\hbar
(i-1)\right)\delta _{1-i,q_{\alpha }-q^f_\beta},\nonumber\\
Z_{\Gamma ,\text{vortex}}^{\text{antifund},q^{af}_\beta}(a,m,\hbar ;k)&=&\prod _{\alpha =1}^N\prod _{\beta=1}^F\prod _{s\in Y_{\alpha }}\left(a_{\alpha }+m_\beta+\hbar
(i-1)\right)\delta _{1-i,q_{\alpha }-q^{af}_\beta}.
\eea

Orbifold vortex partition functions also have many sectors determined by discrete charges:
\be
Z_{\text{vortex}}\left(\left\{a,m,q_{\alpha }^{(L)}\right\};\left\{q_{\alpha }^f\right\}\right)=\sum _k \prod _{\beta =1}^Nz_{\beta }^{k_{\beta }}Z_{\Gamma ,\text{vortex}}^{\text{fund},q_{\beta }^f}(a,m,k)Z_{\Gamma ,\text{vortex}}^{\text{vector}}\left(a,k;\left\{q_{\alpha }\right\}\right),
\label{orbifoldvortex}
\ee
where $z_{\beta}$ are $N$ counting parameters, which are related but not identical to the counting parameters in (8).

\section{Vortex From Instantons}

We can extract orbifold vortex partition functions from orbifold instanton partition functions following similar strategy for non-orbifold case \cite{BTZ2}. Generally speaking, counting parameters of instantons will be combined to give counting parameters of vortices and two dimensional Young tableaux in instanton counting will collapse in a nice way to one dimensional Young tableaux in vortex counting. For $SU(N)$ vortex, we need to consider $SU(N)$ N-node linear quiver gauge theory. The instanton partition function of this gauge theory is characterized by N N-dimensional arrows of Young-tableaux, which in noted by $Y^{(L)}_\alpha$ in (4).
Then by setting masses of antifundamental hypermultiplets and bifundamental hypermultiplets to special values, the Young-tableaux are forced to have following simple form:

\be
Y_{\alpha }^{(L)}=\left\{
\begin{array}{cc}
 k_L & \alpha = L   \\
\emptyset & \text{other} \text{wise}
\end{array}
 \right..
\label{confYoung}
\ee
The readers should keep in mind of the $\delta$-functions of discrete charges which means that not all of the boxes in above Young tableaux will contribute to the partition functions.
Through direct calculation, we will show how to get this constraint naturally. Then we prove the equality between this degenerate orbifold instanton partition function and the $SU(N)$ orbifold vortex partition function. A necessary tool to achieve these goals is the following proposition.

\newtheorem{prop}{Proposition}[section]
\begin{prop}
For generic orbifold space, when $m^{(L)}_{\alpha,\beta}=0$, $Y^{(L)}_\alpha$ should equal to $Y^{(L+1)}_\beta$ and when $m^{(L)}_{\alpha,\beta}=\epsilon_2$ , $Y^{(L+1)}_\beta$ should have one more row than that of $Y^{(L)}_\alpha$. In this latter situation, if we further suppose the orbifold space is $\mathbb{C}/\mathbb{Z}_2$, $Y_{\alpha}^{(L)}$ has $M$ rows with lengths $k_1 \leq k_2 \leq \ldots\leq k_M$ and $Y^{(L+1)}_{\beta}$ had $M+1$ rows with lengths $l_0 \leq l_1 \leq \ldots \leq l_M$, then for $1\leq i \leq M$ either $k_i=l_{i-1}$ or $k_i=l_i+1$.
\label{prop}
\end{prop}

One important observation is that in the self-dual limit \(\epsilon _+=0\), the boxes contribute to orbifold instanton partition function are picked out by their relative hook length. So, upto some modifications the proof of the degeneracy phenomenon in \cite{BTZ2} is valid for the orbifold case and the above proposition can be proved analogously.

\subsection{Constraint from fundamental hypermultiplets}

From the formula (\ref{inst_fund}), we know that for antifundamental hypermultiplets, if we want to get \(Y_{\alpha }=\emptyset\), it is necessary that  \(a_{\alpha }+m_f=0\) and  the box \((1,1)\)
satisfy the $\delta $-function, that is  $q_{\alpha }-q_{\beta }^{\text{af}}=0 \bmod p$ for some $\beta$. On the other hand , if we want to reduce \(Y_{\alpha }\) to be one row, then  \(a_{\alpha }+m_f=-\epsilon _2\) and the box $(1,2)$ should satisfy the $\delta $-function, that is \(q_{\alpha }-q^{af}_\beta=1 \bmod p\) for some $\beta$. In order to satisfy (\ref{confYoung}), we should take:
\be
\left\{
\begin{array}{c}
 a_{\alpha }^{(1)}+m_{\alpha }=- \epsilon _2\text{ } \delta _{\alpha ,1} \\
 q_{\alpha }^{(1)}-q_{\alpha }^{\text{af}}=\delta _{\alpha ,1}\text{  }\text{mod } p
\end{array}
\label{fundpara}
\right..
\ee

\subsection{Constraint from bifundamental hypermultiplets}

Using the proposition \ref{prop}, It is easy to find that in order to satisfy (\ref{confYoung}), following identities should be satisfied:
\be
\left\{
\begin{array}{c}
 m_{\alpha ,\alpha }^{(L)}=\epsilon _2 \delta _{\alpha ,L+1} \\
 Q_{\alpha ,\alpha }^{(L)}= -\delta _{\alpha ,L+1} \bmod p
\end{array}
\label{bifundpara}
\right.,
\ee

which means:

\be
m_{\alpha ,\beta }^{(L)}=\left\{
\begin{array}{cc}
 a_{\alpha ,\beta }^{(L)}=a_{\alpha ,\beta }^{(L+1)} & \alpha \in [1,L]; \beta =[1,L] \nonumber\\
 a_{\alpha ,\beta }^{(L+1)} & a\in [1,L]; \beta \in [L+1,N] \\
 a_{\alpha ,\beta }^{(L)} & \alpha \in [L+1,N]; \beta =[1,L] \nonumber
\end{array}
\right.,
\ee
and
\be
Q_{\alpha ,\beta }^{(L)}=\left\{
\begin{array}{cc}
 q_{\alpha ,\beta }^{(L)}=q_{\alpha ,\beta }^{(L+1)} & \alpha \in [1,L]; \beta =[1,L] \nonumber\\
 q_{\alpha ,\beta }^{(L+1)} & a\in [1,L]; \beta \in [L+1,N] \\
 q_{\alpha ,\beta }^{(L)} & \alpha \in [L+1,N]; \beta =[1,L] \nonumber
\end{array}
\right..
\ee
We see that the pattern of \(Q_{\alpha ,\beta }^{(L)}\) is the same as that of \(m_{\alpha ,\beta }^{(L)}\). This is a necessary consistent condition to extract orbifold vortex partition functions from orbifold instanton partition functions. The following subsection contains technical details of this statement.

\subsection{Reshuffling the partition function}

In order to make formulas lighter, we will make the $\delta $-functions of discrete charges implicit and use following notations:
\bea
\begin{array}{ccc}
 (x)_k^+\text{:=}(x)_k=\prod _{i=0}^{k-1}\left(x+i \epsilon _2\right), & \text{ }  & (x)_k^-\text{:=}\prod _{i=1}^k\left(x-\epsilon _2 i\right).
\end{array}\nonumber
\eea
Now let's input (\ref{bifundpara}) into (8) and find the contribution from the $L$-th vector-multiplet is:
\bea
{\left(Z_L^{\Gamma }\right)}^{-1}&=&A\cdot B\cdot C\nonumber,\\
A&=&\prod _{\alpha ,\beta =1}^L\left(a_{\alpha ,\beta }^{(L)}\right){}_{k_{\alpha },k_{\beta }}\nonumber,\\
B&=&\prod _{\alpha =1}^L\prod _{\beta =L+1}^N(-1)^{k_{\alpha }}\left(a_{\beta ,\alpha }^{(L)}\right){}_{k_{\alpha }}=\prod _{\alpha =1}^L\prod _{\beta
=L+1}^N\left(a_{\alpha ,\beta }^{(L)}\right)_{k_{\alpha }}^- ,\\
C&=&\prod _{\beta =1}^L\prod _{\alpha =L+1}^N\left(a_{\alpha ,\beta }^{(L)}\right)_{k_{\beta }}^+\nonumber.
\eea
After suitable reshuffling we also get the contribution from the $L$-th bifundamental hypermultiplet as:
\bea
Z_{L,L+1}^{\Gamma }&=&\text{I}\cdot \text{II}\cdot \text{III}\nonumber,\\
\text{I}&=&\left\{\prod _{\alpha =1}^L\prod _{\beta =1}^L\left(m_{\alpha ,\beta }^{(L)}\right){}_{k_{\alpha },k_{\beta }}\right\}\nonumber,\\
\text{II}&=&\left\{\prod _{\alpha =1}^L\prod _{\beta =L+2}^N\left(m_{\alpha ,\beta }^{(L)}\right)_{k_{\alpha }}^-\right\}\left\{\prod _{\alpha =1}^L\left(m_{\alpha
,L+1}^{(L)}\right){}_{k_{\alpha },k_{L+1}}\right\},\\
\text{III}&=&\left\{\prod _{\alpha =L+1}^N\prod _{\beta =1}^L\left(m_{\alpha ,\beta }^{(L)}\right)_{k_{\beta }}^+\right\}\left\{\prod _{\alpha =L+1}^N\left(m_{\alpha
,L+1}^{(L)}\right)_{k_{L+1}}^+\right\}\nonumber,
\eea

so:
\be
Z_L^{\Gamma }Z_{L,L+1}^{\Gamma }=\frac{\left\{\prod _{\alpha =1}^L\left(a_{\alpha ,L+1}^{(L+1)}\right){}_{k_{\alpha },k_{L+1}}\right\}}{\left\{\prod _{\alpha =1}^L\left(a_{\alpha
,L+1}^{(L)}\right)_{k_{\alpha }}^-\right\}}\left\{\prod _{\alpha =L+2}^N\left(a_{\alpha ,L+1}^{(L+1)}\right)_{k_{L+1}}^+\right\}\left(\epsilon _2\right)_{k_{L+1}}^+ .
\ee
Other factors are:
\bea
Z_{\Gamma }^{\text{fund}}&=&\prod _{f=1}^N\prod _{i=1}^{k_1}\left(a_1^{(1)}+m_f-\epsilon _2(i-1)\right)=\left(-\epsilon _2\right)_{k_1}^-\prod _{f=2}^N\left(a_{1,f}^{(1)}\right)_{k_1}^-,\\
Z^\Gamma_N&=&\prod _{\alpha =1}^N\left(\epsilon _2\right)_{k_{\alpha }}^+\left(\epsilon _2\right)_{k_{\alpha }}^-\prod _{\alpha <\beta }^N\left(a_{\alpha
,\beta }^{(N)}\right){}_{k_{\alpha },k_{\beta }}\left(a_{\beta ,\alpha }^{(N)}\right){}_{k_{\beta },k_{\alpha }}.
\eea

Parameters in above formulas are not independent, since from the explicit form of \(m_{\alpha ,\alpha }^{(L)}\), we know:

\be
a_{\alpha ,\beta }^{(L+1)}-a_{\alpha ,\beta }^{(L)}=-\epsilon _2 \left(\delta _{\alpha ,L+1}-\delta _{\beta ,L+1}\right)\nonumber.
\ee

It follows that:
\be
\begin{array}{cc}
 a_{K,L}^{(L)}=a_{K,L}^{(N)} & L\in [2,N],K<L ,\\
 a_{K,L+1}^{(L)}=a_{K,L+1}^{(K)} & L\in [K,N-1],K\in [1,N-1].
\end{array}
\ee
Similar relations are found for discrete charges:
\be
\begin{array}{cc}
 Q_{K,L}^{(L)}=Q_{K,L}^{(N)} & L\in [2,N],K<L ,\\
 Q_{K,L+1}^{(L)}=Q_{K,L+1}^{(K)} & L\in [K,N-1],K\in [1,N-1].
\end{array}
\ee
This induce the identification of following factors:
\bea
\prod _{L=1}^{N-1}\left\{\prod _{\alpha =1}^L\left(a_{\alpha ,L+1}^{(L+1)}\right){}_{k_{\alpha },k_{L+1}}\right\}=\prod _{\alpha <\beta }^N\left(a_{\beta
,\alpha }^{(N)}\right){}_{k_{\beta },k_{\alpha }}\left(a_{\alpha ,\beta }^{(N)}\right){}_{k_{\alpha },k_{\beta }}\nonumber,\\
\left\{\prod _{f=2}^N\left(a_{1,f}^{(1)}\right){}_{k_1}\right\}\left\{\prod _{L=1}^{N-1}\prod _{\alpha =L+2}^N\left(a_{L+1,\alpha }^{(L+1)}\right){}_{k_{L+1}}\right\}=\prod
_{L=1}^{N-1}\left\{\prod _{\alpha =1}^L\left(a_{\alpha ,L+1}^{(L)}\right){}_{k_{\alpha }}\right\}\nonumber.
\eea

With these identities we have :

\bea
Z_{\text{Quiver}}(k)=\frac{\prod _{\beta =1}^N Z_{\Gamma }^{\text{fund},q_{\beta }^f}\left(Y^{(N)}\right) }{\prod _{\alpha
=1}^N\left(\epsilon _2\right)_{k_{\alpha }}\prod _{\alpha <\beta }^N\left(a_{\beta ,\alpha }^{(N)}\right){}_{k_{\beta },k_{\alpha }}}.
\eea
The equality in above formula is exact upto an overall sign factor which will disappear after redefine counting parameters. We recognize that the formula above is the same as the orbifold vortex partition function, if we identify $a^{(N)}_\alpha$  and $\epsilon_2$ in (8) with $a_\alpha$ and $\hbar$ in (\ref{orbifoldvortex}). A comment here is that the moduli space of orbifold instanton may have nontrivial first Chern class. We will concentrate on the case when the first Chern class is trivial which will give extra constraints on Young-tableaux. But this does not affect all the arguments in this section.

\section{Vortex on \(\mathbb{C}\left/\mathbb{Z}_2\right.\) and \(\mathcal{N}=1\) Super Liouville Theory}

In \cite{superLiouvilleNS} and \cite{superLiouville}  people discussed about AGT like relation between instanton partition functions on \(\mathbb{C}^2/\mathbb{Z}_2\) and \(\mathcal{N}=1\)
super Liouville theory. In the following, we will study the relation between \(\text{SU}(2)\) vortex on \(\mathbb{C}\left/\mathbb{Z}_2\right.\) and
degenerate states in \(\mathcal{N}=1\) super Liouville theory.

\subsection{SU(2) Vortex on \(\mathbb{C}\left/\mathbb{Z}_2\right.\)}

In order to compare orbifold vortex partition functions with conformal blocks of the $\mathcal{N}=1$ super Liouville theory, it is convenient to rewrite vortex partition functions as linear differential operators acting on products of hypergeometric functions.

\subsubsection*{Vector field contribution }

\bea
\left(Z_{\Gamma ,\text{vortex}}^{\text{vector}}\left(a,\hbar ;k;q_{1,2}\right)\right){}^{-1}&=&U_{\Gamma ,\text{vortex}}^{\text{vector}}(\hbar ,k)O_{\Gamma
,\text{vortex}}^{\text{vector}}\left(a,\hbar ;k;q_{1,2}\right),\\
U_{\Gamma ,\text{vortex}}^{\text{vector}}(\hbar ,k)&=&\prod _{\alpha =1}^2(2\hbar )^{\left\lfloor \frac{k_{\alpha }}{2}\right\rfloor }\left\lfloor \frac{k_{\alpha }}{2}\right\rfloor !\nonumber,\\
O_{\Gamma ,\text{vortex}}^{\text{vector}}\left(a,\hbar ;k;q_{1,2}\right)&=&\prod _{i=1}^{k_1}\left(a_{1,2}+\hbar \left(k_2+1-i\right)\right)\delta
_{-i+1+k_2,q_{1,2}}\nonumber\\&&
\prod _{j=1}^{k_2}\left(a_{2,1}+\hbar \left(k_1+1-j\right)\right)\delta _{-j+1+k_1,q_{2,1}}\nonumber.
\eea
where  $\lfloor x\rfloor$ is the floor function that is the largest integer not greater than $x$. The first part in above formula is an abelian factor. By abelian, we mean that it is the same as corresponding part of abelian vortex partition functions. The second part can be considered
as the essential factor in nonabelian vortex theories. The contributions to the partition functions from vector fields are classified by \(q_{1,2}\). Since \(q_{1,2}\) takes
value in \(\mathbb{Z}_2\), there are two different branches. In the following, we will set \(a_1=a, a_2=-a\) and rewrite the second part as:

\be
O_{\Gamma ,\text{vortex}}^{\text{vector}}(a,\hbar ;k;0)= D_{k_1,k_2}^0\prod _{i=1}^{\left\lfloor \frac{k_1}{2}\right\rfloor }(-2a+2\hbar  i)(2\hbar
 i)\prod _{i=1}^{\left\lfloor \frac{k_2}{2}\right\rfloor }(2a+2\hbar  i)(2\hbar  i),
 \ee
 \be
O_{\Gamma ,\text{vortex}}^{\text{vector}}(a,\hbar ;k;1)=D_{k_1,k_2}^1\prod _{i=1}^{\left\lceil \frac{k_1}{2}\right\rceil }(-2a+\hbar (2 i-1))\prod
_{i=1}^{\left\lceil \frac{k_2}{2}\right\rceil }(2a+\hbar (2 i-1)),
\ee

where the pre-factors are defined as:
\be
D_{k_1,k_2}^0\text{:=}\left\{
\begin{array}{cc}
 \frac{2a (-1)^{\frac{k_1+k_2}{2}}}{2a+\hbar \left(k_2-k_1\right)}(-1)^{k_1} & k_1+k_2 \text{even} \\
 (-1)^{\frac{k_2+1+k_1}{2}}2a & k_1+k_2 \text{odd}
\end{array}
\right.,
\ee
\be
D_{k_1,k_2}^1=\left\{
\begin{array}{cc}
 (-1)^{\frac{k_1+k_2}{2}}(-1)^{k_1} & k_1+k_2 \text{even} \\
 \frac{(-1)^{\frac{k_2-1+k_1}{2}}}{2a+\hbar \left(k_2-k_1\right)}(-1)^{k_1} & k_1+k_2 \text{odd}
\end{array}
\right..
\ee

These pre-factors will turn out to be linear differential operators acting on orbifold vortex partition functions.

\subsubsection*{Fundamental hypermultiplets contribution}

Since \(q_{\alpha }-q_f\) can only take values of 0 and 1, there are four type contributions from fundamental hypermultiplets.
When $q_{1,2}=0$, we have:
\bea
Z_{\Gamma ,\text{vortex}}^{\text{fund},0,0}\left(a,m_f,\hbar ;k\right)&=&\prod _{\alpha =1}^2\prod _{i=1}^{\left\lceil \frac{k_{\alpha }}{2}\right\rceil }\left(m_{\alpha ,f}+2\hbar (i-1)\right),\\
Z_{\Gamma ,\text{vortex}}^{\text{fund},0,1}(a,m,\hbar ;k)&=&\prod _{\alpha =1}^2\prod _{i=1}^{\left\lfloor \frac{k_{\alpha }}{2}\right\rfloor }\left(m_{\alpha ,f}+\hbar (2i-1)\right)\nonumber,
\eea
where $m_{\alpha ,f}=a_{\alpha }-m_f$ and $\lceil x\rceil$ is the ceiling  function that is the smallest integer not less than x. When $q_{1,2}=1$, we have:
\bea
Z_{\Gamma ,\text{vortex}}^{\text{fund},1,0}\left(a,m_f,\hbar ;k\right)&=&\prod _{i=1}^{\left\lceil \frac{k_1}{2}\right\rceil }\left(m_{1,f}+2\hbar (i-1)\right)\prod _{i=1}^{\left\lfloor \frac{k_2}{2}\right\rfloor }\left(m_{2,f}+\hbar (2i-1)\right),\\
Z_{\Gamma ,\text{vortex}}^{\text{fund},1,1}(a,m,\hbar ;k)&=&\prod _{i=1}^{\left\lfloor \frac{k_1}{2}\right\rfloor }\left(m_{1,f}+\hbar (2i-1)\right)\prod _{i=1}^{\left\lceil \frac{k_2}{2}\right\rceil }\left(m_{2,f}+2\hbar (i-1)\right)\nonumber.
\eea
Notice that on the LHS of the formula above we use two integers in the superscript to denote the types of fundamental hypermultiplet contributions.

\subsubsection*{Vortex partition functions}

Unlike non-orbifold case, where there is only one vortex partition function, orbifold vortex partition function has many sectors characterized by discrete charges.
\bea
Z_{\Gamma }^{\text{vortex}}\left(q_{1,2},p_1,p_2;k\right)&\text{:=}&Z_{\Gamma ,\text{vortex}}^{\text{vector}}\left(a,\hbar ;k;q_{1,2}\right)Z_{\Gamma ,\text{vortex}}^{\text{fund},q_{1,2},p_1}\left(a,m_1,\hbar ;k\right)\\
&&Z_{\Gamma ,\text{vortex}}^{\text{fund},q_{1,2},p_2}\left(a,m_2,\hbar ;k\right)\nonumber.
\eea
On the LHS of above formula we make $a$ and the mass parameters implicit to make the formula shorter. In general there are eight different types, since the integers of the LHS can only take values in 0 and 1. Four examples related to our discussion are:
\be
Z_{\Gamma }^{\text{vortex}}(0, 0,0;k)=\frac{1}{D_{k_1,k_2}^0}\frac{\prod _{\alpha =1}^2\prod _{i=1}^{\left\lceil \frac{k_{\alpha }}{2}\right\rceil }\left(m_{\alpha ,1}+2\hbar (i-1)\right)\prod _{i=1}^{\left\lceil \frac{k_{\alpha }}{2}\right\rceil }\left(m_{\alpha ,2}+2\hbar (i-1)\right)}{ \prod _{i=1}^{\left\lfloor \frac{k_1}{2}\right\rfloor }(-2a+2\hbar  i)(2\hbar  i)\prod _{i=1}^{\left\lfloor \frac{k_2}{2}\right\rfloor }(2a+2\hbar  i)(2\hbar  i)}
\label{vortex1},
\ee
\be
Z_{\Gamma }^{\text{vortex}}(0, 0,1;k)=\frac{1}{D_{k_1,k_2}^0}\frac{\prod _{\alpha =1}^2\prod _{i=1}^{\left\lceil \frac{k_{\alpha }}{2}\right\rceil }\left(m_{\alpha ,1}+2\hbar (i-1)\right)\prod _{i=1}^{\left\lfloor \frac{k_{\alpha }}{2}\right\rfloor }\left(m_{\alpha ,2}+2\hbar (i-1)\right)}{ \prod _{i=1}^{\left\lfloor \frac{k_1}{2}\right\rfloor }(-2a+2\hbar  i)(2\hbar  i)\prod _{i=1}^{\left\lfloor \frac{k_2}{2}\right\rfloor }(2a+2\hbar  i)(2\hbar  i)}
\label{vortex2},
\ee
\be
Z_{\Gamma }^{\text{vortex}}(1, 0,0;k)=\frac{1}{D_{k_1,k_2}^1}\frac{\prod _f^2\prod _{i=1}^{\left\lceil \frac{k_1}{2}\right\rceil }\left(m_{1,f}+2\hbar (i-1)\right)\prod _{i=1}^{\left\lfloor \frac{k_2}{2}\right\rfloor }\left(m_{2,f}+\hbar (2i-1)\right)}{ \prod _{\alpha =1}^2(2\hbar )^{\left\lfloor \frac{k_{\alpha }}{2}\right\rfloor }\left\lfloor \frac{k_{\alpha }}{2}\right\rfloor ! \prod _{i=1}^{\left\lceil \frac{k_1}{2}\right\rceil }(-2a+\hbar (2 i-1))\prod _{i=1}^{\left\lceil \frac{k_2}{2}\right\rceil }(2a+\hbar (2 i-1))}
\label{vortex3},
\ee
\be
Z_{\Gamma }^{\text{vortex}}(1, 0,1;k)=\frac{1}{D_{k_1,k_2}^0}\frac{\prod _f^2\prod _{i=1}^{\left\lfloor \frac{k_1}{2}\right\rfloor }\left(m_{1,f}+\hbar (2i-1)\right)\prod _{i=1}^{\left\lceil \frac{k_2}{2}\right\rceil }\left(m_{2,f}+2\hbar (i-1)\right)}{\prod _{\alpha =1}^2(2\hbar )^{\left\lfloor \frac{k_{\alpha }}{2}\right\rfloor }\left\lfloor \frac{k_{\alpha }}{2}\right\rfloor ! \prod _{i=1}^{\left\lceil \frac{k_1}{2}\right\rceil }(-2a+\hbar (2 i-1))\prod _{i=1}^{\left\lceil \frac{k_2}{2}\right\rceil }(2a+\hbar (2 i-1))}
\label{vortex4}.
\ee

Since there are more branches of orbifold instanton partition functions than the types of four point correlation functions, it is reasonable that not all kinds of orbifold instanton partition function has a super Liouville theory explanation. Correspondingly not all of above vortex partition functions will correspond to correlation functions with degenerate states in super Liouville theory. Considering the symmetry between fundamental and antifundamental hypermultiplets of linear quiver gauge theories, we will show in following subsections only (\ref{vortex2}), (\ref{vortex3}), and (\ref{vortex4}) may have conformal filed theory explanations.  Let's first concentrate on (\ref{vortex2}):

\bea
&&Z_{\Gamma }^{\text{vortex}}\left(0, 0,1\right)\text{:=}\sum _k z_1^{k_1}z_2^{k_2}Z_{\Gamma }^{\text{vortex}}\left(0, 0,1;k\right)\text{:=}\nonumber\\
&&\sum _l\left(z_1^{2l_1}z_2^{2l_2}Z_{\Gamma }^{\text{vortex}}\left(0, 0,1;\left\{2l_1,2l_2\right\}\right)+z_1^{2l_1}z_2^{2l_2+1}Z_{\Gamma
}^{\text{vortex}}\left(0, 0,1;\left\{2l_1,2l_2+1\right\}\right)\right.\nonumber\\
&&\left.+z_1^{2l_1+1}z_2^{2l_2}Z_{\Gamma }^{\text{vortex}}\left(0, 0,1;\left\{2l_1+1,2l_2\right\}\right)+z_1^{2l_1+1}z_2^{2l_2+1}Z_{\Gamma
}^{\text{vortex}}\left(0, 0,1;\left\{2l_1+1,2l_2+1\right\}\right)\right)\nonumber,
\eea

where $l_1$ and  $l_2$  are non-negative integers.

For $l_1$ and $l_2$ even:
\bea
&&\sum _lz_1^{2l_1}z_2^{2l_2}Z_{\Gamma }^{\text{vortex}}\left(0, 0,1;\left\{2l_1,2l_2\right\}\right)=\\
&&\left(1+\frac{\hbar }{2a}\left(z_2\partial _{z_2}-z_1\partial _{z_1}\right)\right)F\left(\frac{m_{1,1}}{2\hbar },\frac{m_{1,2}}{2\hbar },\frac{-2a}{2\hbar }+1,-z_1^2\right)F\left(\frac{m_{2,1}}{2\hbar },\frac{m_{2,2}}{2\hbar },\frac{2a}{2\hbar }+1,-z_2^2\right)\nonumber.
\eea
For $l_1$ even and $l_2$ odd:
\bea
&&\sum _lz_1^{2l_1}z_2^{2l_2+1}Z_{\Gamma }^{\text{vortex}}\left(0, 0,1;\left\{2l_1,2l_2+1\right\}\right)=\\
&&\frac{-z_2}{2a}F\left(\frac{m_{1,1}}{2\hbar },\frac{m_{1,2}}{2\hbar },\frac{-2a}{2\hbar }+1,-z_1^2\right)F\left(\frac{m_{2,1}}{2\hbar },\frac{m_{2,2}}{2\hbar },\frac{2a}{2\hbar }+1,-z_2^2\right)\nonumber.
\eea
For $l_1$ odd and $l_2$ even:
\bea
&&
\sum _lz_1^{2l_1+1}z_2^{2l_2}Z_{\Gamma }^{\text{vortex}}\left(0, 0,1;\left\{2l_1+1,2l_2\right\}\right)=\\
&&-\frac{m_{1,1}m_{1,2}z_1}{2a}F\left(\frac{m_{1,1}}{2\hbar }+1,\frac{m_{1,2}}{2\hbar }+1,\frac{-2a}{2\hbar }+1,-z_1^2\right)F\left(\frac{m_{1,1}}{2\hbar },\frac{m_{2,1}}{2\hbar },\frac{2a}{2\hbar }+1,-z_2^2\right)\nonumber.
\eea
For $l_1$ odd and $l_2$ odd:
\bea
&&\sum _lz_1^{2l_1+1}z_2^{2l_2+1}Z_{\Gamma }^{\text{vortex}}\left(0, 0,1;\left\{2l_1+1,2l_2+1\right\}\right)=\\
&&
z_1z_2m_{1,1}m_{1,2}\left(1+\frac{\hbar }{2a}\left(z_2\partial _{z_2}-z_1\partial _{z_1}\right)\right)F\left(\frac{m_{1,1}}{2\hbar }+1,\frac{m_{1,2}}{2\hbar }+1,\frac{-2a}{2\hbar }+1,-z_1^2\right)\nonumber\\
&&
F\left(\frac{m_{2,1}}{2\hbar },\frac{m_{2,2}}{2\hbar },\frac{2a}{2\hbar }+1,-z_2^2\right)\nonumber.
\eea

Separately, each of them can be considered as some intertwine differential operators acting on products of two hypergeometric functions.

Another type of vortex partition function which we want to calculate explicitly is (\ref{vortex4}):
\bea
&&
Z_{\Gamma }^{\text{vortex}}\left(1, 0,1\right)\text{:=}\sum _kz_1^{k_1}z_2^{k_2}Z_{\Gamma }^{\text{vortex}}\left(q_{1,2}=1, 0,1;k\right)\text{:=}\\
&&
\sum _l\left(z_1^{2l_1}z_2^{2l_2}Z_{\Gamma }^{\text{vortex}}\left(1, 0,1;\left\{2l_1,2l_2\right\}\right)+z_1^{2l_1}z_2^{2l_2+1}Z_{\Gamma
}^{\text{vortex}}\left(1, 0,1;\left\{2l_1,2l_2+1\right\}\right)\right.\nonumber\\
&&
\left.+z_1^{2l_1+1}z_2^{2l_2}Z_{\Gamma }^{\text{vortex}}\left(1, 0,1;\left\{2l_1+1,2l_2\right\}\right)+z_1^{2l_1+1}z_2^{2l_2+1}Z_{\Gamma
}^{\text{vortex}}\left(1, 0,1;\left\{2l_1+1,2l_2+1\right\}\right)\right)\nonumber.
\eea
For $l_1$ even and $l_2$ even:
\bea
&&
\sum _l z_1^{2l_1}z_2^{2l_2}Z_{\Gamma }^{\text{vortex}}\left(1, 0,1;\left\{2l_1,2l_2\right\}\right)=\nonumber\\
&&
F\left(\frac{m_{1,1}}{2\hbar },\frac{m_{1,2}}{2\hbar },\frac{-2a}{2\hbar }-\frac{3}{2},-z_1^2\right)F\left(\frac{m_{2,1}}{2\hbar },\frac{m_{2,2}}{2\hbar },\frac{2a}{2\hbar }-\frac{3}{2},-z_2^2\right).
\eea
For $l_1$ even and $l_2$ odd:
\bea
&&
\sum _lz_1^{2l_1}z_2^{2l_2+1}Z_{\Gamma }^{\text{vortex}}\left(1, 0,1;\left\{2l_1,2l_2+1\right\}\right)=z_2\frac{2a+2\hbar +\hbar \left(z_2\partial _{z_2}-z_1\partial _{z_1}\right)}{2a+\hbar }\nonumber\\
&&
F\left(\frac{m_{1,1}}{2\hbar },\frac{m_{1,2}}{2\hbar },\frac{-2a}{2\hbar }-\frac{3}{2},-z_1^2\right)F\left(\frac{m_{2,1}}{2\hbar },\frac{m_{2,2}}{2\hbar },\frac{2a}{2\hbar }+\frac{1}{2},-z_1^2\right).
\eea
For $l_1$ odd and $l_2$ even:
\bea
&&
\sum _lz_1^{2l_1+1}z_2^{2l_2}Z_{\Gamma }^{\text{vortex}}\left(1, 0,1;\left\{2l_1+1,2l_2\right\}\right)=z_1m_{1,1}m_{1,2}\frac{2a-2\hbar +\hbar \left(z_2\partial _{z_2}-z_1\partial _{z_1}\right)}{2a-\hbar }\nonumber\\
&&
F\left(\frac{m_{1,1}}{2\hbar }+1,\frac{m_{1,2}}{2\hbar }+1,\frac{-2a}{2\hbar }+\frac{3}{2},-z_1^2\right)F\left(\frac{m_{2,1}}{2\hbar },\frac{m_{2,2}}{2\hbar },\frac{2a}{2\hbar }+\frac{1}{2},-z_2^2\right).
\eea
For $l_1$ odd and $l_2$ odd:
\bea
&&
\sum _lz_1^{2l_1+1}z_2^{2l_2+1}Z_{\Gamma }^{\text{vortex}}\left(1, 0,1;\left\{2l_1+1,2l_2+1\right\}\right)=\frac{z_1z_2m_{1,1}m_{1,2}}{(-2a+\hbar )(2a+\hbar )}\nonumber\\
&&
F\left(\frac{m_{1,1}}{2\hbar }+1,\frac{m_{1,2}}{2\hbar }+1,\frac{-2a}{2\hbar }+\frac{3}{2},-z_1^2\right)F\left(\frac{m_{2,1}}{2\hbar },\frac{m_{2,2}}{2\hbar },\frac{2a}{2\hbar }+\frac{3}{2},-z_2^2\right).
\eea
A universal property of \(\text{SU}(2)\) \(\mathbb{Z}_2\) orbifold vortex partition functions is that they are quadratic forms of Gaussian hypergeometric
functions. This is the same for non-orbifold case and one big difference is the effective counting parameter is \(2\hbar\) for orbifold case while
\(\hbar\) for non-orbifold case. We will see the CFT correspondence of these properties.

\subsection{Relation to super Liouville theory}

Since we know the relation between orbifold vortex partition function and orbifold instanton partition function, we can find the relation between
orbifold and vortex through degeneration procedure on super Liouville theory side. Recall that, $SU(N)$ vortex partition functions come from $SU(N)$ quiver gauge theory with N nodes. We are now interested in
$SU(2)$ gauge theory with two nodes, and therefore we have five points on a sphere. There are in principle two ways. (1) Calculate directly the correlation
function between two lowest degenerate states and three non-degenerate primary states in \(\mathcal{N}=1\) super Liouville theory. (2) If we know
the complete AGT relation between partition functions of SU(2) instantons on \(\mathbb{C}^2/\mathbb{Z}_2\) and correlation functions of \(\mathcal{N}=1\)
super Liouville theory with both Ramond and NS primary fields, we get the relation between orbifold vortex and \(\mathcal{N}=1\) super Liouville
theory almost for free. However, technically both ways are difficult. There are no results concerning (1) and (2) in the literature. In the following we will use existing results to analysis the AGT dual of orbifold vortices.

\subsubsection*{Correlation functions with degenerate fields}

As it is clear from previous calculation, in order to extract vortex partition functions from instanton partition functions, the parameters \(m_{\alpha
,\alpha }^{(L)}\) should restrict to be $0$ or \(\epsilon _2\). This means on the CFT side the fusion rule is that from lowest degenerate states, i.e. those
with momentum equals \(\frac{-b}{2}\). It is known that the lowest degenerate states in NS- and R-sector  have momentum equal to \(\frac{-3b}{2}\) and
\(\frac{-b}{2}\) respectively. So the CFT dual of $SU(2)$ orbifold vortex should come from five point correlation functions with two lowest degenerate
states in the R-sector.


\begin{figure}
\begin{center}
\includegraphics[ width=0.8\textwidth]{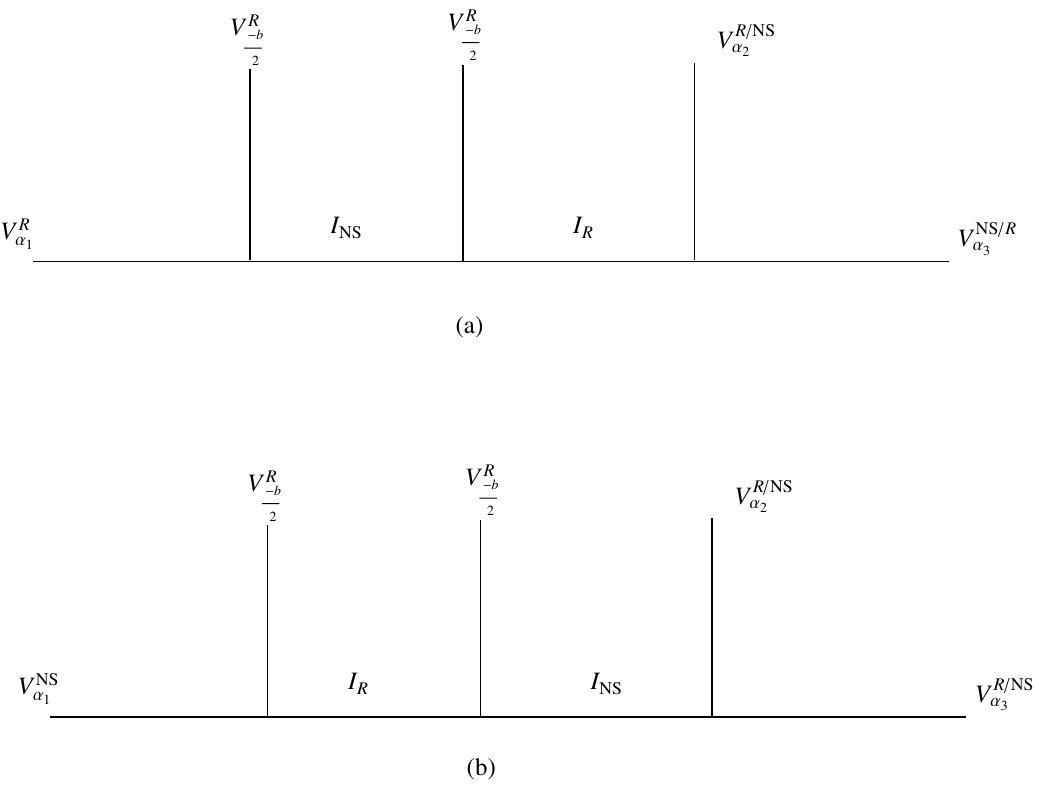}
\caption{five point correlation functions corresponding to $SU(2)$ $\mathbb{Z}_2$ orbifold vertices.}
\label{fig5pt}
\end{center}
\end{figure}
Possible configurations are show in Fig \ref{fig5pt}, where $V_{\alpha }^R$ and $V_{\alpha }^{NS}$ denote  primary fields with momentum $\alpha$ in Ramond- and NS-sector and
$I_R, I_{NS}$ are identity operators in Ramond- and NS-sector. To exactly check our proposal, we need to know the AGT correspondence of the following correlation functions in the super Liouville theory:
\bea
\begin{array}{ccc}
 \left\langle V_{\alpha _1}^RV_{\alpha _2}^RV_{\alpha _3}^RV_{\alpha _4}^R\right\rangle {}_{\text{NS}} & \text{and} & \left\langle V_{\alpha _1}^{\text{NS}}V_{\alpha _2}^RV_{\alpha _3}^RV_{\alpha _4}^{\text{NS}}\right\rangle {}_R,
 \label{4pt1}
\end{array}
\\
\begin{array}{ccc}
 \left\langle V_{\alpha _1}^{\text{NS}}V_{\alpha _2}^RV_{\alpha _3}^{\text{NS}}V_{\alpha _4}^R\right\rangle {}_R & \text{and}  & \left\langle V_{\alpha _1}^RV_{\alpha _2}^RV_{\alpha _3}^{\text{NS}}V_{\alpha _4}^{\text{NS}}\right\rangle {}_{\text{NS}}.
 \label{4pt2}
\end{array}
\eea
The subscripts in above correlation functions are used to emphasize the types of internal states. Notice that except the first correlation function in (\ref{4pt1}), the other three are four point correlation function with two Ramond and two NS primary fields. The latter three are not trivially related, since they have different internal states.

The first internal state of the correlation function in Fig \ref{fig5pt}.a is in NS sector and correspondingly the Kac determinant which gives denominators of conformal blocks is also in NS
sector. From \cite{superLiouvilleNS}, we can expect that \(q_1^{(1)}=q_2^{(1)} \bmod 2\), since they determine the form of denominators of
instanton partition functions (\ref{inst:vector}).  Similarly,  from \cite{Ito}, we will conjecture that \(q_1^{(2)}=q_2^{(2)}+1 \bmod 2\).  According
to \cite{BTZ2}, the fusion rule of the first  \(V_{\frac{-b}{2}}^R\) corresponds to the choice (\ref{fundpara}),  this means that  when \(q_1^{(1)}=q_2^{(1)}
\bmod 2\), we have \(q_1^{\text{af}}=q_2^{\text{af}}+1 \bmod 2\) and when \(q_1^{(1)}=q_2^{(1)} +1 \bmod 2\), we have \(q_1^{\text{af}}=q_2^{\text{af}}\text{
 }\text{mod} 2\).  Our choice of the discrete charges is different from that of \cite{superLiouvilleNS}, which in our language is \(q_1^{(1)}=q_2^{(1)}
\bmod 2\)  and  \(q_1^{\text{af}}=q_2^{\text{af}}\text{  }\text{mod} 2\). If we further  consider the symmetry between fundamental and antifundamental
hypermultiplets,  \(q_{\alpha }^f=q_{\alpha }^{\text{af}}\), we find that only (\ref{vortex2}), (\ref{vortex3}), (\ref{vortex4}) can be identified as correlation
function in Fig \ref{fig5pt}.a

\begin{figure}
\begin{center}
\includegraphics[ width=0.8\textwidth]{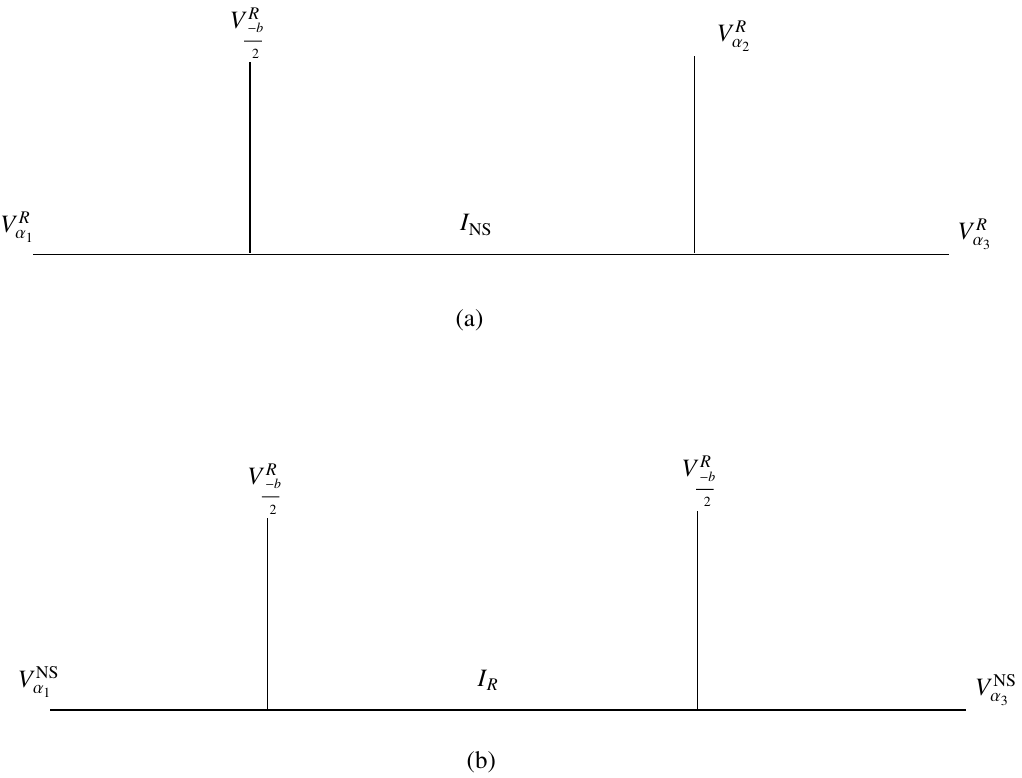}
\caption{four point correlation functions in Super Liouville theory}
\label{fig4pt}
\end{center}
\end{figure}

However, presently there are no results in the literature of super Liouville theory that we can use to give a direct check of our claim. What we know are the four point correlation functions in Fig.\ref{fig4pt}, which are calculated in \cite{hep-th/0202032}.

For $\left\langle V_{\alpha _1}^{NS} V_{\frac{-b}{2}}^RV_{\alpha _2}^RV_{\alpha _3}^{NS}\right\rangle$, the hypergeometric function factors are:
\bea
&&F\left(\frac{1}{2b^{-1}}\left(\alpha _1+\alpha _3+\alpha _4\right)+\frac{3}{4},\frac{1}{2b^{-1}}\left(a_1+\alpha _3-\alpha _4\right)+\frac{3}{4},\frac{2\alpha _1}{2b^{-1}}+\frac{3}{2}\right),\\
&&F\left(\frac{1}{2b^{-1}}\left(a_1+\alpha _3+\alpha _4\right)+\frac{1}{4},\frac{1}{2b^{-1}}\left(a_1+\alpha _3-\alpha _4\right)+\frac{1}{4},\frac{2\alpha _1}{2b^{-1}}+\frac{1}{2}\right).
\eea


For $\left\langle V_{\alpha _1}^RV_{\frac{-b}{2}}^RV_{\alpha _2}^RV_{\alpha _3}^R\right\rangle$, the hypergeometric function factors are:

\bea
&&F\left(\frac{1}{2b^{-1}}\left(a_1+\alpha _3+\alpha _4\right)+\frac{3}{4},\frac{1}{2b^{-1}}\left(a_1+\alpha _2-\alpha _3\right)+\frac{3}{2},\frac{2\alpha _1}{2b^{-1}}+\frac{3}{2}\right),\\
&&
F\left(\frac{1}{2b^{-1}}\left(a_1+\alpha _3+\alpha _4\right)+\frac{3}{4},\frac{1}{2b^{-1}}\left(a_1+\alpha _2-\alpha _3\right)+\frac{3}{2},\frac{2\alpha _1}{2b^{-1}}+\frac{1}{2}\right).
\eea

The one for Fig \ref{fig4pt}.b is also calculated in \cite{Chorazkiewicz2011} with a different convention,
\be
F\left(\frac{1}{2b^{-1}}\left(\alpha _1+\alpha _2-\alpha _3\right)-\frac{1}{4}, \frac{1}{2b^{-1 }}\left(\alpha _1+\alpha _2+\alpha _3\right)-\frac{1}{4},\frac{2\alpha _1}{2b^{-1}}+1\right).
\ee

We can see that after a linear map between parameters of orbifold vortices and degenerate four point correlation functions in super Liouville theory, we can identify the hypergeometric function factors of both sides.
\bea
b^{-1}&=&\hbar\nonumber,\\
   \alpha_1&=&a+\text{const}\nonumber,\\
\alpha _2+\alpha _3&=&m_1+\text{const}\nonumber,\\
\alpha _2+\alpha _3&=&m_2+\text{const}\nonumber.
\eea

The constants depends on which pair of hypergeometric functions we are comparing. This is an evidence that orbifold vortex partition functions should correspond to correlation functions of lowest degenerate Ramond fields as show in Fig.\ref{fig5pt}. It also tells us that the identification of parameters of orbifold instanton partition functions and that of correlation functions of the super Liouville theory in mixed sectors is the same--upto a constant shift--as in original AGT paper \cite{AGT}.

It is important to notice that as in non-orbifold case the four point correlation functions in Fig.\ref{fig4pt} can not be identified with Abelian vortex partition function, since the former has three parameters-- the three momentums, while the latter has only two parameters--the two masses of fundamental hypermultiplets. So a direct check of our proposal should start from a direct clear calculation of the correlation functions in Fig.\ref{fig5pt}, which is a hard problem due to the subtleties coming from the multi-branch of super conformal generator in R-sector and also the double  vacuua in R-sector. We leave this problem in future study.

If we consider four point correlation functions with one degenerate fields as the ``partition'' function of surface operators, we will have two types of simple surface operators in the gauge theory dual of $\mathcal{N}=1$ super Liouville theory, since super Liouville theory has two types of lowest degenerate states. Exactly, for $\mathbb{Z}_2$ orbifold $SU(2)$ gauge theory  with flavor number equals 2, the instanton partition functions only  have two types of lowest degeneration.

\section{Discussions}

We consider some functions which are the four dimensional limit of strip amplitudes satisfying the same $\delta$-functions of discrete charges as orbifold instanton partition functions and denote them by $ \mathcal{A}(a,m,Y)$, where $a$ and $m$ are parameters associated with Coulomb branch parameters and masses, and Y are N-dimensional arrow of Young-tableaux. Then a natural question is whether we can reduce orbifold instanton partition functions of a quiver gauge theory to these functions with general Young-tableaux as we did for non-orbifold case \cite{BTZ2}. By the proposition \ref{prop}, we can show that it is doable for two situations. (1)If discrete charges take value in $\mathbb{Z}_p$ for general $p$, $Y$ should be an arrow of N rows, which is just the vortex case. (2) If discrete charges take value in $\mathbb{Z}_2$, $Y$ can be arbitrary. This makes the $\mathbb{Z}_2$ case especially simple and it is  expected to interpret simple surface operators in $\mathbb{Z}_2$ orbifold gauge theory as degenerate fields in $\mathcal{N}=1$ super Liouville theory.

Using degenerate fields as a probe, we should be able to get a full AGT correspondence between instanton partition functions on $\mathbb{C}/\mathbb{Z}_2$ and $\mathcal{N}=1$ super Liouville theory. Exactly, we get a relation between a certain branch of instanton partition functions and the correlation function with four primary Ramond fields and check this relation up-to three instanton contributions. Further checks to higher order instanton contributions and other types of correlation functions are left for future work \cite{toappear}.

\section*{ Acknowledgements}
The author sincerely thanks  G.~Bonelli, K.~Maruyoshi and A.~Tanzini for valuable discussions and comments and V.~Belavin, L.~Hadasz, P.~Suchanek for patience in answering his questions about super Liouville theory.
J.Z. is partially supported by the INFN project TV12.

\end{document}